\documentclass[amsmath,amssymb,aps,prl,reprint,floatfix]{revtex4-2}

\usepackage{amssymb}
\usepackage{amsmath}
\usepackage[usenames,dvipsnames]{xcolor}
\usepackage[colorlinks=true,citecolor=blue,linkcolor=blue]{hyperref}
\usepackage{graphicx}
\usepackage{float}
\usepackage{booktabs} 
\usepackage{makecell}
\usepackage{titlesec}
\usepackage{dcolumn}
\usepackage{bm}
\usepackage{physics}
\usepackage{siunitx}
\usepackage{hyperref}
\hypersetup{
    colorlinks=true,
    citecolor=blue,
    linkcolor=blue,
    urlcolor=blue
}
\usepackage{colortbl,xcolor}

\titlespacing{\subsection}{0pt}{12pt}{6pt}

\makeatletter
\newcommand{\relUnc}[2]{$#1\times10^{#2}$}
\newcommand{\Sr}{Sr$^+$}
\newcommand{\Ca}{Ca$^+$}

\makeatother

\begin{document}
\title{Direct comparison of multi-ion optical clocks based on $^{40}$Ca$^+$ and $^{88}$Sr$^+$}

\author{Yosef Sokolik}
\affiliation{Department of Physics of Complex Systems and AMOS, Weizmann Institute of Science,
Rehovot 7610001, Israel}

\author{Roee Ozeri}
\affiliation{Department of Physics of Complex Systems and AMOS, Weizmann Institute of Science,
Rehovot 7610001, Israel}

\author{Nitzan Akerman}
\affiliation{Department of Physics of Complex Systems and AMOS, Weizmann Institute of Science, Rehovot 7610001, Israel}

\begin{abstract}
We report the first direct frequency comparison between two multi-ion optical clocks based on the S$_{1/2}$ to D$_{5/2}$ transition in \Ca and \Sr ions.  Using linear chains of up to nine \Ca ions and six \Sr ions, we demonstrate improved stability as a function of the number of ions that are contributing to the laser frequency stabilization servo. The measured joint fractional frequency stability of the two clocks reaches \relUnc{1.37(12)}{-15} at one second, placing an upper bound on the same stability of one of the clocks at \relUnc{9.6(8)}{-16} in one second. We measured the frequency ratio of the two clocks to be $R_{\text{Sr/Ca}}=1.082076536381896986(18)$, where the systematic uncertainty is primarily limited by the room temperature blackbody radiation. Our direct measurement represents an order of magnitude improvement compared to existing indirect frequency ratio measurements. Furthermore, by combining our results with recent absolute frequency measurements of the \Sr transition, referenced to a primary frequency standard, we refined the absolute frequency of the \Ca transition to $\nu_{\text{Ca}^+}=411042129776400.21(4)$ Hz, reducing its uncertainty by a factor of three. This study presents the first direct comparison between two multi-ion optical clocks, highlighting their significant potential for future applications in fundamental physics tests, geodesy, and precision metrology.

\end{abstract}

\maketitle

Optical atomic clocks have reached fractional frequency uncertainties at the $10^{-18}$ level \cite{collaboration2021frequency, Hausser2025In}, marking significant progress for precision metrology with applications ranging from timekeeping to fundamental physics \cite{safronova2018search, Chu2025Exploring,mcgrew2018atomic,takamoto2020test,sanner2019optical}. The exceptional accuracy of atomic clocks arrives at the cost of quantum projection noise (QPN), an intrinsic source of statistical uncertainty that limits the stability of single-atom measurements. Achieving state-of-the-art stability near the $10^{-18}$ level using a single atom requires long averaging times, making these systems less suitable for applications that require detecting fast time-varying phenomena\cite{Lange2021Improved,roberts2020search}. However, increasing the number of atoms allows for faster averaging of projection noise, improving stability by a factor of $\sqrt{N}$ according to the standard quantum limit (SQL). This collective measurement principle underlies the development of optical lattice\cite{aeppli2024clock} and tweezers\cite{young2020half} clocks, which exploit ensembles of neutral atoms to enhance stability.

Extending this strategy to trapped-ion systems introduces new challenges. Multi-ion spectroscopy suffers from inhomogeneous frequency shifts, caused by electric and magnetic field variations within the traps. Two prominent examples are the linear Zeeman shift (LZS)  and the quadrupole shift (QS) \cite{itano2000external}. While some of these effects can be effectively canceled in single-ion systems, their impact grows with increasing ion number, making traditional methods inadequate. Recent demonstrations of multi-ion clocks have addressed this issue through various strategies, including using multi-zone single-ion trapping \cite{morrison2025autonomous}, selecting atomic species with reduced sensitivity to field gradients \cite{keller2019controlling}, operating near the quadrupole-shift nulling angle \cite{tan2019suppressing}, or dynamically reorienting the magnetic field \cite{lange2020coherent}. Among these, dynamical decoupling techniques—where multiple Zeeman components are coherently averaged, have proven particularly promising \cite{kaewuam2020hyperfine,pelzer2024multi,akerman2025operating}.

Multi-ion optical clocks offer enhanced stability through increased ion numbers, provided systematic frequency shifts can be precisely controlled and suppressed. In a previous work \cite{akerman2025operating}, we demonstrated the value of using dynamical decoupling in the operation of trapped-ion optical clocks by self-comparison in a single clock. Our approach utilizes dynamic decoupling sequences to effectively mitigate the LZS and QS \cite{Shaniv2019QuadrupoleDecoupling}, which arise from magnetic field inhomogeneities and ion trapping potentials. Additionally, we employ common techniques to mitigate other systematic shifts. For example, we performed micromotion sideband spectroscopy to minimize excess micromotion, with residual shifts mitigated using a magic RF trap frequency to balance the ac-Stark shift and second-order Doppler shift \cite{dube2014high,huang2019ca+}.

Using the above techniques, we present the first direct comparison between two multi-ion clocks based on $^{40}$\Ca and $^{88}$\Sr, each exploiting the S$_{1/2}$ to D$_{5/2}$ quadrupole transition. We study the stability of the clocks with linear crystals of up to nine ions, showing the enhancement in performance as a function of the number of ions. We then use the two clocks to directly measure, for the first time, the frequency ratio between the two species' clock transitions. Our result reduces the uncertainty of this ratio by an order of magnitude, as compared with previous indirect measurements. Our result is mostly limited by the uncertainty of black-body radiation (BBR) shift at the level of \relUnc{2}{-17}. We note that a significantly tighter evaluation of BBR ($<$\relUnc{1}{-18}) has been previously demonstrated in room temperature clocks \cite{lindvall202588} and can be implemented in our system in future work.
 
\begin{figure}[t!]
    \centering
    \includegraphics[trim={6cm 5.4cm 7cm 3cm},clip,width=1\linewidth]{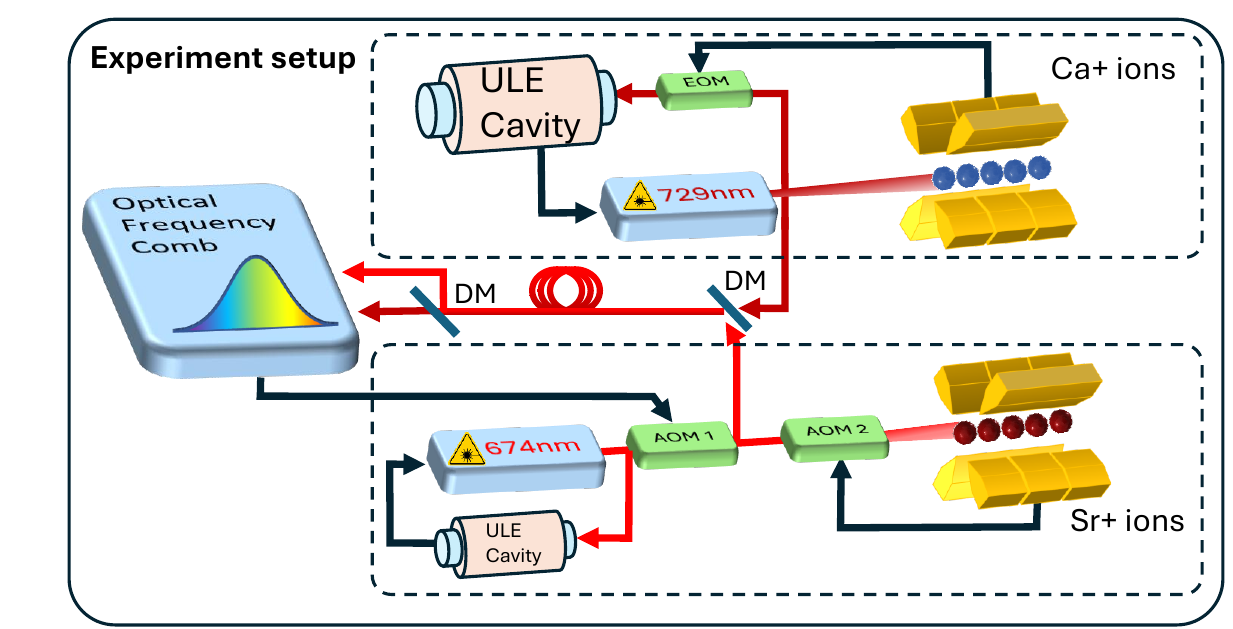}
    \caption{Experimental setup for the direct comparison of multi-ion Ca$^+$ and Sr$^+$ optical clocks. The 729~nm Ca$^+$ clock laser is pre-stabilized to a ULE cavity and locked to the Ca$^+$ $S_{1/2}\!\rightarrow\!D_{5/2}$ transition via an EOM offset. It stabilizes an OFC, which transfers coherence to the pre-stabilized 674~nm Sr$^+$ clock laser. The light of the two lasers is delivered to the OFC through a common fiber to suppress fiber-induced phase noise. No active stabilization of fiber noise is used in the setup.}
    \label{system}
\end{figure} 
Our experimental setup consists of two similar but separate apparatuses, one for \Ca ions and the other for \Sr ions. Each consists of a compact vacuum chamber with a single-zone segmented linear Paul trap, where an in-vacuum current-carrying wire is used to drive RF transitions between different Zeeman levels. A single layer of magnetic shielding surrounds the chamber, allowing us to achieve Zeeman coherence on the order of a second, which is significantly longer than the coherence of the optical clock.  

Figure \ref{system} presents a schematic diagram of the setup used in our clocks comparison. The \Ca clock laser at 729 nm is pre-stabilized to a high-finesse ULE cavity with short-term stability of \relUnc{\sim1}{-15} between 1-10 seconds. An EOM is used to offset (in addition to generating the PDH signal) the laser frequency and bridge the difference between the atomic transition and the cavity resonance. We lock the laser to the \Ca clock transition by adjusting this offset frequency. The 729 laser also stabilizes an optical frequency comb (OFC) to which the \Sr clock laser at 674 nm is locked. The 674 nm is pre-stabilized to another ULE cavity, however, its relevant spectral characteristics are inherited from the OFC.

Optical fibers are used to deliver light between different components of the setup, specifically from the lasers to the ion traps and to the OFC. None of the fibers is actively compensated, however, they are passively isolated to reduce phase noise. While the fibers to the traps are kept short ($<$3m), the length to OFC is around 10m. Since the two lasers are positioned on the same optical table, we transfer their light to the OFC by the same fiber by combining them on a dichroic mirror. This way, the effect of fiber phase noise, which is common to both lasers, is suppressed.  

We are using a quasi-continuous dynamical decoupling (QCDD) sequence to interrogate the clock transition. It is a Ramsey-like spectroscopy scheme with the addition of a resonant RF drive during the Ramsey time, which coherently averages the Zeeman state in both ground and excited manifolds in a way that cancels the LZS and QS. This interrogation sequence is described in detail in ref \cite{Shaniv2019QuadrupoleDecoupling,akerman2025operating}. 

For the \Ca clock, the Ramsey time accounts for 70\% of the entire acquisition time, including all SPAM and experiment control. For detection, we use an FPGA to read and analyze the camera images in real-time (using the available CameraLink output of the iXon987 camera), and correct for the laser detuning after each measurement frame with a non-blocking latency of around 50ms (limited by the control electronics of the EOM frequency source). 

\begin{figure}[t]
    \centering
    \includegraphics[trim={0cm 0cm 0cm 1cm},clip,width=1\linewidth]{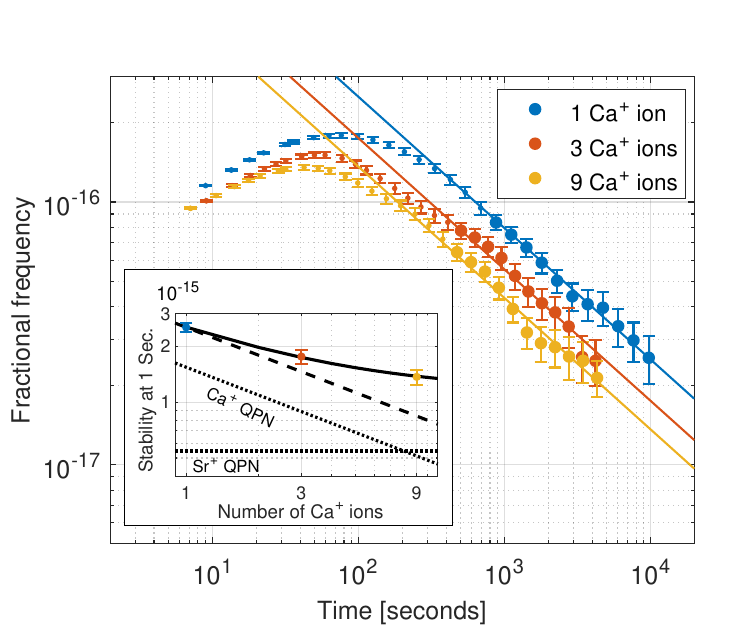}
    \caption{Clock stability vs. number of ions. The main figure presents an overlapping Allan deviation of the frequency ratio of two clocks for three different numbers of \Ca ions used in the laser servo, and a fixed number of six \Sr ions. Solid lines represent linear fits on a log-log scale, with data points at longer time scales (indicated by large markers). The inset shows our extracted stability at one second, from the fits, as a function of the number of \Ca ions. The dashed black line is a fit to $1/\sqrt{N}$, and the solid black line is a fit that includes residual laser noise. The dotted lines indicate the calculated pure QPN of the two clocks.}
    \label{Stability}
\end{figure}

\section{Multi-ion clock stability} 
Using a linear crystal consisting of nine \Ca ions, we study the stability of the multi-ion \Ca clock while varying the number of ions used in the analysis and for operating the laser frequency servo. The performance is assessed by comparison to the \Sr clock, operated with six ions. Figure~\ref{Stability} shows an overlapping Allan-deviation analysis of the \Sr clock in three cases, in which the 729-nm laser (which locks the 674-nm laser through the OFC) is servoed using one (blue), three (red), or nine (yellow) \Ca ions. The Ramsey time for the \Ca clock is $\tau_{R,\text{Ca}}=100$ms, and for \Sr it is $\tau_{R,\text{Sr}}=200$ms.

The solid lines are linear fits on a log-log scale to the measured data points at averaging times $> 300$s, where the servo behavior has a reduced impact on the stability. In the inset, we plot the extrapolated stability at one second, obtained from those fits, as a function of the number of \Ca ions. The dashed black line indicates the expected $\sqrt{N}$ scaling anchored to the single-ion value, where the dotted line represents the QPN-limited stability of the two clocks. We observe that the stability improves with increasing ion number but deviates from a $\sqrt{N}$ scaling. This behavior is consistent with a regime in which the laser instability  $\sigma_L(\tau_R)$ starts to dominate over QPN at larger $N$.

Assuming white laser noise, we empirically model the measured stability as $\sigma_c(\tau=1s)=\sqrt{\sigma_L^2+(\sigma_{N}/\sqrt{N})^2}$. Fitting this model to the data (solid black line) yields $\sigma_L=$\relUnc{1.1(3)}{-15} and $\sigma_{N}=$\relUnc{2.2(4)}{-15}. We vary the number of \Ca ions rather than the number of \Sr ions because the vacuum conditions in the \Sr system impede reliable operation with more than six ions. We note that the servo gain used to stabilize the 729-nm laser to the ions was adjusted according to the ion number to prevent phase slips, since the SNR also varies with $N$.

Using nine \Ca ions, we obtain a two-clock comparison stability of \relUnc{1.37(12)}{-15} at one second. This sets an upper bound on the stability of one of the clocks to \relUnc{9.6(8)}{-16} at one second, which places this result not far from the most stable ion clocks \cite{marshall2025high,dietze2025entanglement,kim2023improved}, despite starting with only moderate ($\sim$ \relUnc{1}{-15} at one second) pre-stabilized laser performance.

\section{Frequency ratio of \texorpdfstring{$^{40}$Ca$^+$}{40Ca+} and \texorpdfstring{$^{88}$Sr$^+$}{88Sr+} clock transitions}
\begin{figure}[t]
    \centering
    \includegraphics[trim={0cm 0.25cm 0.5cm 0cm},clip,width=1\linewidth]{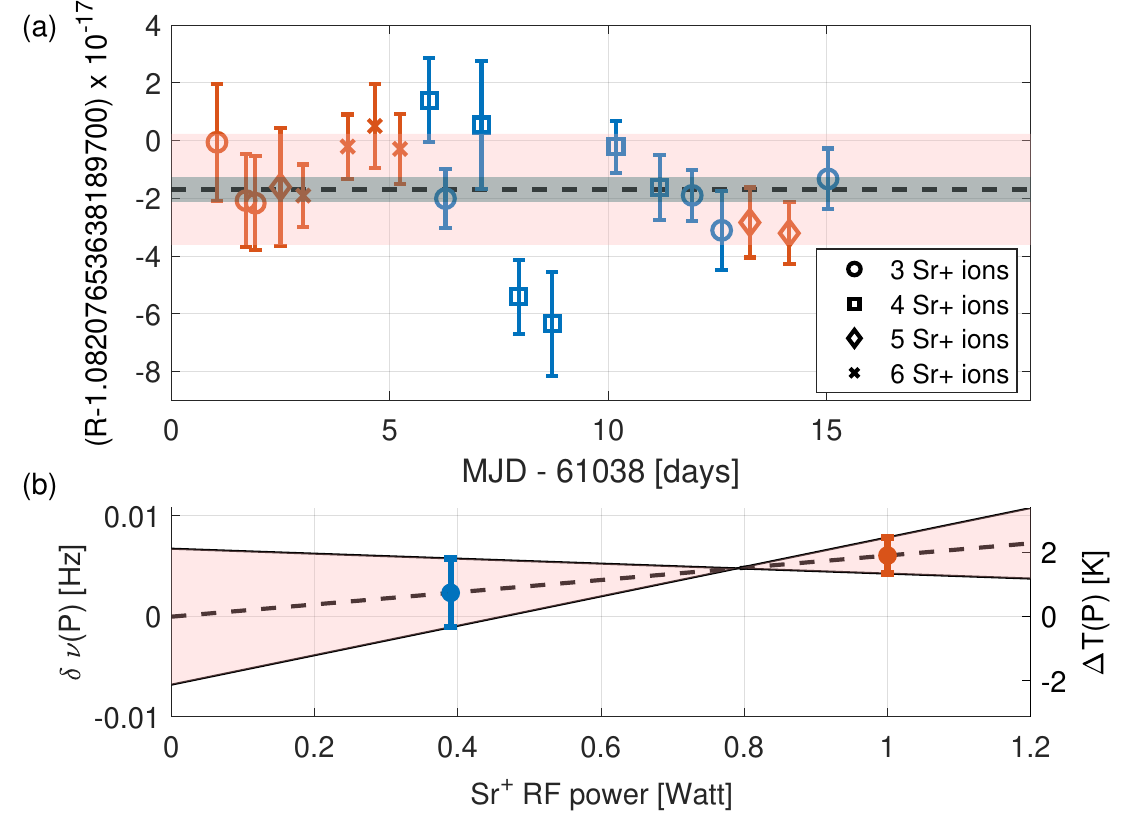}
    \caption{Frequency ratio of the $^{88}$Sr$^+$ and $^{40}$Ca$^+$ clock transitions.
            vhv fh;;;;(a) Individual measurement runs corrected for all the systematic effects collected over two weeks. Symbols indicate the number of \Sr ions that were used in the measurement. The color corresponds to the two different trap RF powers for which the \Sr clock was operated, as presented in (b). The mean, standard deviation, and statistical uncertainty of the mean are indicated by the dashed line, red shaded area, and gray shaded area, respectively. 
            (b) Frequency shift of the \Sr clock, extracted from the measured ratio, as a function of trap RF drive power due to the BBR beyond the measured chamber temperature. The linear fit and uncertainty are indicated by the dashed black line and red shaded area, respectively. We used this analysis to correct for the BBR shift in the ratio results of (a).}
    \label{ratio}
\end{figure} 

We now describe our direct measurement of the frequency ratio between the two clock transitions. Using the same methodology as described above, we repeatedly measure the frequency ratio over two weeks under various conditions (e.g., different ion numbers and trap RF drive strengths), with a total acquisition time of 170 hours. Our results are shown in Fig.~\ref{ratio}(a). Each data point corresponds to a single measurement run lasting a few hours, with a typical statistical uncertainty of about \relUnc{1-3}{-17}. The different symbols indicate the number of \Sr ions used in the comparison, while the number of \Ca ions was fixed to eight. Combining all results, we obtain $R_{\text{Sr/Ca}}=1.082076536381896986(18)$. The relative uncertainty of \relUnc{18}{-18} is dominated by systematic effects, as the statistical uncertainty of \relUnc{4}{-18} is well below it. The systematic uncertainty evaluation is summarized in Table~\ref{table}. In the following, we briefly discuss the dominant contributions, where a more detailed analysis is provided in the supplementary material \cite{sm}.

\begin{table}[b]
\caption{\label{tab:SrCaShifts}
Fractional frequency shifts and $1\sigma$ uncertainties for the
${}^{88}\mathrm{Sr^+}$ and ${}^{40}\mathrm{Ca^+}$ systems, expressed in units of $10^{-18}$.}
\begin{ruledtabular}
\begin{tabular}{lcccc}
\textrm{Effect} &
\multicolumn{2}{c}{${}^{88}\mathrm{Sr^+}$} &
\multicolumn{2}{c}{${}^{40}\mathrm{Ca^+}$} \\
\textrm{} &
\textrm{Shift} &
\textrm{Unc.} &
\textrm{Shift} &
\textrm{Unc.} \\
\colrule
Quadratic Zeeman                            & $686$   & $0.7$   & $2432$   & $3 $    \\
Blackbody radiation (BBR)          & $\rm (P_L)529$   & $5$     & $891$    & $12$   \\
                                    & $\rm(P_H)538$   & $14$    &          &   \\
Secular motion (2$^{\mathrm{nd}}$ Doppler)  & $-0.8$  & $0.3$   & $-4.4$   & $2 $  \\
Height difference (redshift)                &         &         & $11$     & $0.5$  \\
Excess Micromotion                          & $0.4$   & $0.2$   & $0.15$    & $0.07$ \\
Quadruple Shift                             & $0$     & $1$     & $0$      & $1$ \\
\colrule
\textbf{Total correction}           &\textbf{1223} & \textbf{14} & \textbf{3329} & \textbf{12} \\
\end{tabular}
\end{ruledtabular}
\label{table}
\end{table}
\vspace{5pt}
\textbf{Blackbody Radiation (BBR).} The dominant contribution to our systematic uncertainty is the BBR shift. At around room
temperature, the susceptibilities of the clock transitions are estimated to be $\Delta\nu^{\text {Sr}}_{BBR} = 3.0616(46) \times 10^{-11} \times T^4$ Hz/K$^4$ \cite{steinel2023evaluation} and $\Delta\nu^{\text {Ca}}_{BBR} = 4.7066(3)\times10^{-11}\times T^4$ Hz/K$^4$ \cite{huang2022liquid}. While we measure the temperature of the vacuum chambers of the clocks to 0.1 K, resistive heating of the trap assembly due to RF currents can increase the effective BBR temperature experienced by the ions. 

Since we have no temperature sensor on the trap assembly, we follow the technique presented in Ref.~\cite{steinel2023evaluation} to constrain the BBR shift. We perform a clock comparison at two different \Sr trap RF drive power $\rm P_L=0.4$ Watt and $\rm P_H=1$ Watt. Our results are shown in Fig.~\ref{ratio}(b). Assuming a linear dependence on RF power, we obtain a frequency shift for the \Sr clock of \relUnc{6(9)}{-3}Hz/Watt around room temperature. Using the above susceptibilities, we convert the shift to an effective temperature difference seen by the ions. We use this result to account for the dependence of the BBR shift on the trap RF power for both clocks, and correct each measurement to obtain the final BBR-corrected ratio presented in Fig.~\ref{ratio}(a). We take a conservative approach to the uncertainty and add the BBR uncertainties of the two clocks, even though these are expected to be correlated and partially cancel in the frequency ratio estimate. This yields BBR relative uncertainty of \relUnc{18}{-18}.

\vspace{5pt}
\textbf{Magnetic Fields.} 
A static bias magnetic field of $B_{\mathrm{dc}}\approx\SI{3.1}{G}$ for the \Sr system and $B_{\mathrm{dc}}\approx\SI{2.6}{G}$ for the \Ca system is applied along the ion crystal axis to define the quantization axis for the laser interactions. The field is generated by permanent magnets inside the shield and fine-tuned with small current coils that compensate for residual gradients along the ion-chain, and correct for long-term drifts in the field magnitude. The relatively high magnetic field that is used here leads to a significant second-order Zeeman shift \cite{itano2000external} ($\sim1$ Hz for the \Ca clock). Since the magnetic field can be measured with high precision, our uncertainty is dominated by the uncertainty in the quadratic coefficient, which has been measured for \Ca \cite{zhang2025liquid} but only calculated for \Sr. We apply the same measured  \relUnc{1}{-3} uncertainty to both coefficients. 

We limit the magnetic field variation to $\approx 10\mu$G during the data acquisition by interleaving Ramsey spectroscopy of the $\mathrm{S_{1/2}}$ manifold and feedback on the coils. The contribution of the second-order ac-Zeeman shift from the trap RF drive is negligible \cite{sm}.
We alternately interrogate the $\left|S_{1/2},m_{J}=\pm1/2\right\rangle\leftrightarrow\left|D_{5/2},m_{J}=\pm3/2\right\rangle$  transitions to cancel the RF-induced shift from the DD pulses in the QCDD scheme \cite{akerman2025operating}. 

\vspace{5pt}
\textbf{Excess Micromotion (EMM).} Excess micromotion (EMM) occurs when the ion is displaced from the RF null \cite{Berkeland1998Micromotion}. This can be a result of stray electric fields, phase difference between electrodes, and, in the multi-ion case, axial micromotion that nulls only at a single point along the ion trap axis, due to the finite size and asymmetry of the trap electrodes. Our traps have an inherent structural asymmetry, as the small gaps between the segmented DC electrodes are not mirrored in the RF electrodes. This configuration prevents the scenario in which misalignment in the assembly will cause homogeneous axial EMM, which can not be compensated for by a DC field. This, however, comes at the price of symmetrically increasing axial EMM from the central ion.   

To evaluate EMM amplitudes, we measure the ratio of the micromotion sideband to the carrier Rabi frequencies in three non-coplanar probe directions. From these ratios, the modulation index and thus the EMM amplitude can be extracted. The Radial EMM can be well compensated in the two radial directions, such that the dominant EMM component is along the axial direction and depends on the position of the ion in the crystal. When the ions are positioned symmetrically around the RF null, the maximally measured modulation index in a 10-ion chain is below $\beta<0.2$.
The EMM causes a negative second-order Doppler shift, and in the case of \Sr and \Ca, a positive scalar quadratic Stark shift. The two effects are canceled at the magic frequencies of 14.35 \cite{dube2014high, Lindvall2025Measurement}MHz and 24.6 MHz\cite{huang2019ca+}, respectively. Here, the \Ca clock operates close to the magic frequency at 24.825 MHz, which largely suppresses the shift due to micromotion. The \Sr trap operates at 16.15 MHz and thus cancellation between the two shifts is smaller. Despite the partial cancellation, since our \Sr clock operates with fewer ions (due to vacuum constraints), the EMM shift in both systems remains below \relUnc{1}{-18} with no significant contribution to the overall uncertainty budget.

\vspace{5pt}
\textbf{Thermal motion.} The motion at the secular frequencies, ~$\omega_i$, produces a second-order Doppler shift. In general, this motion is also accompanied by inherent micromotion. However, as discussed above, magic trap frequency suppresses second-order Doppler due to micromotion, leaving only the contribution of the secular velocity to this shift. 

In the clock interrogation, the ions are cooled to a sub-Doppler temperature using electromagnetically induced transparency (EIT) cooling \cite{lechner2016electromagnetically}, such that all the modes have a phonon occupation of $\bar {n} < 5$. However, during the long Ramsey time, heating occurs, resulting in a linear increase in the number of phonons. It is mostly the center of mass modes that are heated up, as the ambient electric field is largely homogeneous on the scale of the ion crystal. 
The mean phonon occupations, ~$\bar n$, are extracted from the dephasing of carrier Rabi oscillations following Doppler and EIT cooling, and the heating rates~$\dot{\bar n}$ are obtained from linear fits of~$\bar n(t)$ as a function of the delay time. Assuming a linear heating model, the effective occupation averaged over the Ramsey free-evolution time~$\tau_R$ is
$\left<\bar n\right>=n_0 + \dot{\bar n}\cdot\tau_R /2$.

For \Sr the heating rates are $\dot{\bar n}_{z}\approx 2~\mathrm{s^{-1}}$ and $\dot{\bar n}_{x,y}\approx30~\mathrm{s^{-1}}$. For secular frequencies  $\omega_z/2\pi = 596$ kHz and $(\omega_x,\omega_y)/2\pi=(2.241,\,2.275)$ MHz. These low values result in a total shift of order $10^{-18}$ and are insignificant compared with the current experimental precision. The \Ca heating rates are order of magnitude higher, $\dot{\bar n}_{z}\approx30~\mathrm{s^{-1}}$, $\dot{\bar n}_{x}\approx 20~\mathrm{s^{-1}}$ and $\dot{\bar n}_{y}\approx 300~\mathrm{s^{-1}}$. For secular frequencies $\omega_z/2\pi = 474$ kHz and $(\omega_x,\omega_y)/2\pi=(2.63,\,2.69)$ MHz. These values yield a total shift of \relUnc{-5(2)}{-18}, where only one of the radial modes (most likely due to technical noise) dominates the shift.
While the heating rate measurements are accurate to a few percent, the shift uncertainty we state above accounts for possible variation in their values during the clocks' comparisons, as they have been monitored only between runs.  

\vspace{5pt}
\textbf{Electric Quadrupole.} The dynamical decoupling sequence eliminates the electric-quadrupole shift, as well as the second-order tensor Stark shift, owing to their similar operator structure.   The residual frequency shift arising from the non-commuting terms neglected in the QCDD sequence scales inversely with the square of the RF-drive Rabi frequency and is therefore negligible under typical experimental conditions~\cite{Shaniv2019QDD}.  
This ensures that both quadrupole and tensor Stark contributions are suppressed well below the current uncertainty budget of the frequency ratio measurement.

\begin{figure}[t]
    \centering
    \includegraphics[trim={0.5cm 0.2cm 0cm 0cm},clip,width=1\linewidth]{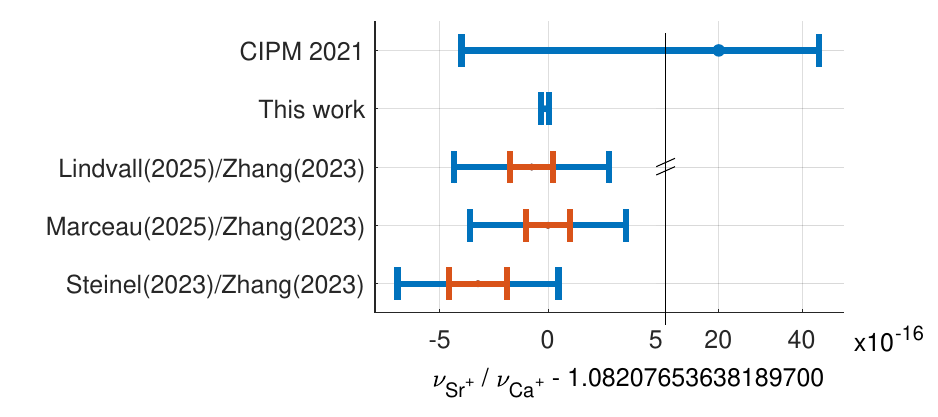}
    \caption{The direct measured frequency ratio from this work compared to indirect combined published values from Zhang(2023)\cite{zhang2023absolute} for \Ca and Lindvall(2025)\cite{lindvall202588}, Marceau(2025)\cite{marceau2025absolute} and Steinel(2023)\cite{steinel2023evaluation} for \Sr. The current CIPM2021\cite{Margolis2024CIPM} recommendation is shown as well. Note the break in the x-axis scale.}
    \label{comp}
\end{figure}

With the systematic error budget at hand, we compare our direct frequency ratio measurement to indirect ratios that are obtained by combining measurements of the absolute transition frequency of the two species. For the \Ca, the most precise result is given by ref \cite{zhang2023absolute}, and for the \Sr, three recent results are used \cite{lindvall202588,marceau2025absolute,steinel2023evaluation}. The comparison is presented in Fig.\ref{comp}. The total uncertainties in the ratios are marked by the blue error bars, where for the indirect ratio, the contribution of the \Sr frequency is indicated by the red error bars.  

While our measurement is in agreement with previous indirect results, it reduces the uncertainty of the ratio by an order of magnitude. Moreover, we can combine our measured ratio with the more accurate \Sr absolute transition frequency to improve on the absolute frequency of the \Ca clock transition. Taking a weighted average of the three \Sr results mentioned here, results in  $\nu_{\text{Ca}^+}=411042129776400.21(4)$ Hz, reducing the uncertainty by a factor of three. 

\section{Summary}

In conclusion, we have carried out a direct frequency comparison between two multi-ion optical clocks based on the secondary frequency standards \Ca and \Sr. The comparison demonstrates enhanced clock stability beyond that of single-ion systems.  This result not only highlights the potential for improving the performance of state-of-the-art ion clocks but also has important implications for applications requiring compact or transportable frequency standards, where the use of ultra-stable cavities is constrained, and for the detection of time-varying signals where short-term clock stability is critical. The successful direct comparison between the \Ca and \Sr multi-ion clocks validates the scalability of the multi-ion approach and establishes a foundation for next-generation optical frequency standards \cite{Dimarcq_2024,lindvall2025coordinated}. It further opens new opportunities in precision metrology, fundamental physics tests, and relativistic geodesy, where higher stability and accuracy can be leveraged for both laboratory and field-deployable systems.

\textit{Acknowledgments -} This work was supported by the Israel Ministry of Science (IMOS Grant No. 3-17376), Israel Science Foundation (ISF Grant No. 1364/24) and by the Lower Saxony (Germany)—Israel Joint Research Program (Grant No. 76251-99-6/20).

\bibliographystyle{apsrev4-2}
\bibliography{references.bib}

\end{document}


\title{ Supplemental Material for:\\ Direct comparison of multi-ion optical clocks based on $^{40}$Ca$^+$ and $^{88}$Sr$^+$}

\author{Yosef Sokolik}
\affiliation{Department of Physics of Complex Systems and AMOS, Weizmann Institute of Science,
Rehovot 7610001, Israel}

\author{Roee Ozeri}
\affiliation{Department of Physics of Complex Systems and AMOS, Weizmann Institute of Science,
Rehovot 7610001, Israel}

\author{Nitzan Akerman}
\affiliation{Department of Physics of Complex Systems and AMOS, Weizmann Institute of Science, Rehovot 7610001, Israel}

\maketitle

\section{Atomic frequency servo implementation}
\label{supp:atomic_servo}
\begin{figure}[t]
\centering
\includegraphics[trim={8cm 4.5cm 8cm 2cm},clip,width=\linewidth]{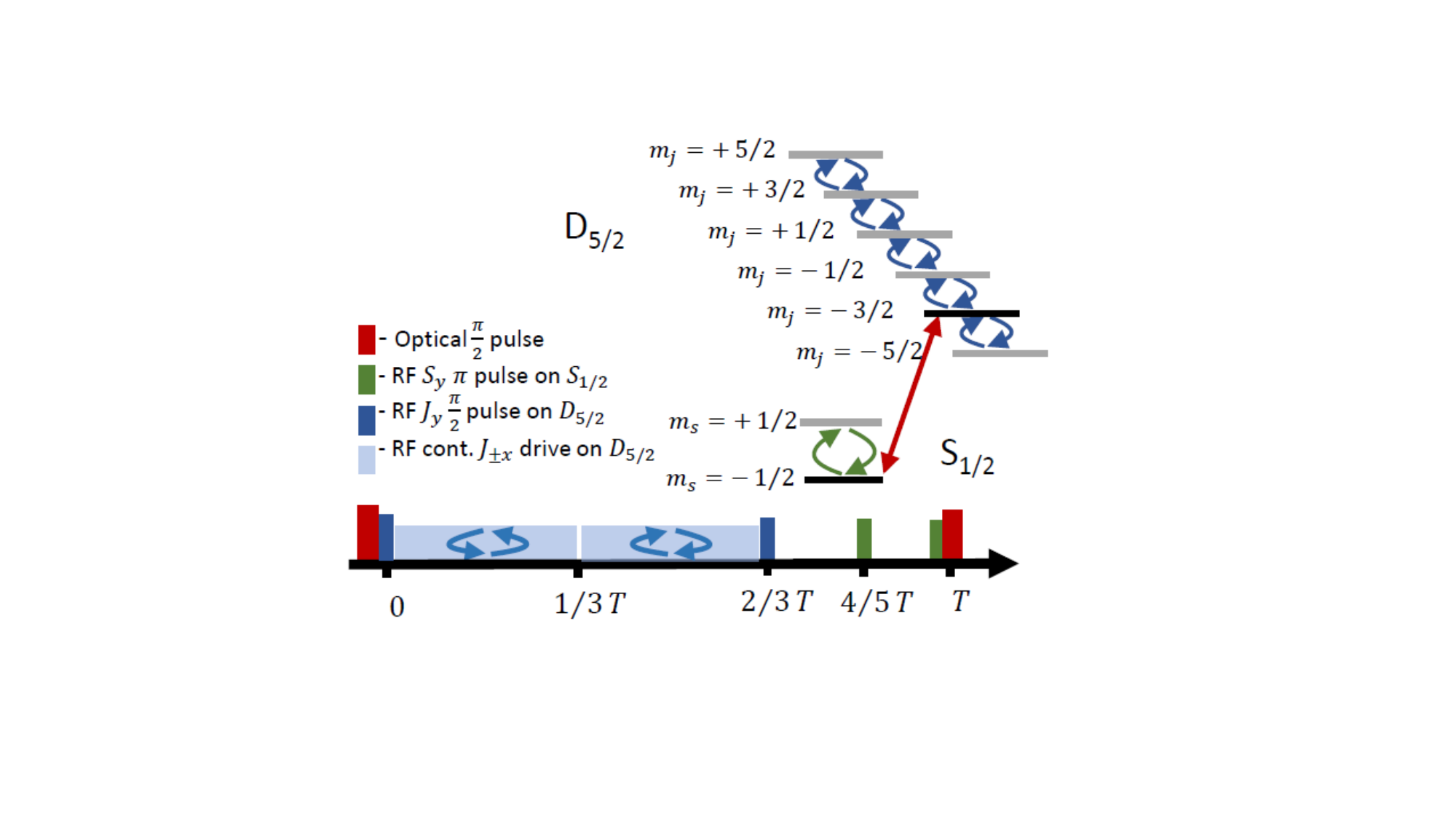}
\caption{The QCDD scheme. A Ramsey-like spectroscopy scheme with
the addition of an RF drive during the Ramsey time. A continuous RF drive (light blue), on resonance with the Zeeman splitting of the $\rm D_{5/2}$ levels, is applied along Jx for the first 2/3 of the Ramsey time. The drive is applied
between two Jy $\pi/2$ pulses, and its sign is flipped in the
middle to unwind the accumulated geometrical phase. The last 1/3 of the Ramsey time consists of only two short $\pi$ pulses (green), which are applied to the S$_{1/2}$ levels to balance the linear Zeeman shift.}
\label{fig:QCDD sequence}
\end{figure}

Our clock interrogation employs a quasi-continuous dynamical decoupling (QCDD) scheme, which combines driving on-resonance RF Zeeman transitions embedded within an optical Ramsey spectroscopy sequence. The sequence cancels the linear Zeeman shift and the electric-quadrupole shift (and concurrently mitigates other tensorial shifts) \cite{Shaniv2019QDD}. Fig~\ref{fig:QCDD sequence} illustrates our full sequence. The resulting excitation signal is identical to that of standard Ramsey spectroscopy and is directly compatible with conventional frequency servo techniques.

The center frequency of each clock transition is stabilized using a four-point QCDD-Ramsey servo, operating on the two optical transitions:
\[
\left|S_{1/2},m_{J}=\pm 1/2\right\rangle \leftrightarrow 
\left|D_{5/2},m_{J}=\pm 3/2\right\rangle
\]
where each Zeeman component is interrogated with two detunings:
\begin{equation}
\delta_{\pm}=\pm\frac{1}{4\tau_{R}},
\end{equation}
corresponding to operation near the half-fringe points of the Ramsey spectrum, where the excitation probability is close to $50\%$.

In the Ca$^{+}$ clock, the two Zeeman components are combined at the level of the interrogation sequence to generate a single clock signal. Successive interrogations alternate between $\delta_{+}$ and $\delta_{-}$, and the resulting deviations from half excitation are combined with alternating sign to form a single dispersive frequency discriminator which is used directly for frequency stabilization.

The Ca$^{+}$ servo stabilizes the 729~nm laser directly, which in turn stabilizes the 674~nm Sr$^{+}$ clock laser via an optical frequency comb. The frequency correction is implemented as a fast feedback loop: the error signal is evaluated at every interrogation cycle, integrated, and applied with a latency of approximately $50~\mathrm{ms}$ by updating the DAC voltage that controls the frequency of an electro-optic modulator (EOM). The EOM provides a tunable frequency offset that bridges the cavity resonance and the Ca$^{+}$ clock transition, and adjusting this offset constitutes the atomic frequency correction.

In the Sr$^{+}$ clock, the two Zeeman components are interrogated independently. For each component, the population imbalance between the $\delta_{\pm}$ interrogations defines a discrete-time frequency error signal, yielding two Zeeman-resolved discriminators. These are subsequently decomposed into a common-mode component, which stems from laser frequency fluctuations, and a differential component, which arises from the initial miscalibration. Both the common-mode and differential corrections are implemented as slow feedback loops (in contrast to the Ca$^+$ servo), as they are not expected to vary in normal clock operations. Frequency error signals are constructed from block-averaged excitation probabilities at $\delta_{\pm}$ for each Zeeman component and averaged over several consecutive clock cycles (typically 3--4) before updating the DDS frequency profiles defining the interrogation frequencies used in the subsequent cycle.

The Crystals' integrity is continuously monitored during servo operation. If de-crystallization is detected, the affected interrogation cycles are discarded, and the servo resumes once stable fluorescence is restored. Collisions with background gas that do not melt the ion crystal do not contribute a significant systematic shift \cite{Hankin2019Systematic} 

\section{Interaction with magnetic fields}

This section provides experimental and theoretical details related to magnetic-field stabilization and magnetic-field-induced frequency shifts, complementing the discussion in the main text.

\subsection{Magnetic-field control and suppression of linear shifts}

\begin{figure}[t]
\centering
\includegraphics[trim={0.5cm 6.5cm 1cm 6.5cm},clip,width=\linewidth]{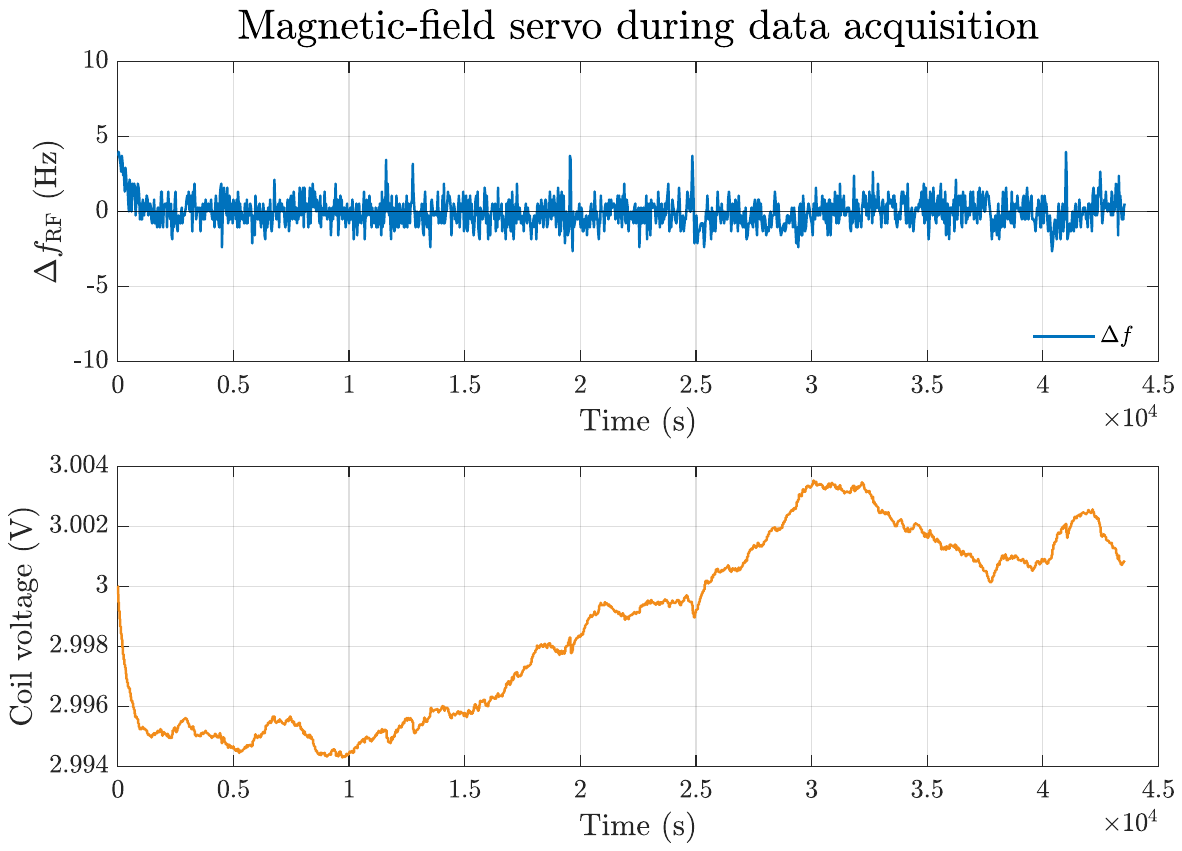}
\caption{Time evolution of the RF-drive detuning (top) and applied compensation-coil voltage (bottom) during operation of the ${}^{88}\mathrm{Sr}^{+}$ clock.}
\label{fig:magnetic field servo}
\end{figure}

A static bias magnetic field defines the quantization axis for clock interrogation. Its magnitude is continuously monitored via spectroscopy of the $\mathrm{S_{1/2}}$ Zeeman splitting and stabilized during clock operation by interleaving Ramsey measurements with slow feedback to compensation coils. This scheme suppresses long-term magnetic-field drifts during data acquisition, limiting field variations to $\lesssim 10~\mu\mathrm{G}$.
Figure~\ref{fig:magnetic field servo} illustrates the performance of the magnetic-field servo for the $\mathrm{Sr}^{+}$ system. The upper plot shows the spectroscopy signal, which is used as the error signal to actively compensate for magnetic-field drifts during clock operation. The lower plot presents the servo output that goes to the compensation coils. Although our QCDD scheme is, in principle, magnetic-insensitive, the magnetic servo is implemented to make sure the DD pulses in the scheme are applied on-resonance with the relevant transitions. Keeping their detuning $\delta<<\Omega_{DD}$ is required for successful suppression of the LZS and QS.

Clock interrogation alternates between the $\left|S_{1/2},m_J=\pm\tfrac{1}{2}\right\rangle \leftrightarrow\left|D_{5/2},m_J=\pm\tfrac{3}{2}\right\rangle$
transitions. Averaging the two Zeeman components produces a composite clock frequency that is insensitive to the frequency of the RF pulses used in the QCDD, and their RF-induced shifts \cite{akerman2025operating}.

\subsection{Quadratic Zeeman shift}

At the bias fields used here, the dominant residual magnetic contribution arises from the quadratic Zeeman effect. In the $\mathrm{D_{5/2}}$ manifold, the second-order shift originates from magnetic-dipole coupling to the nearby $\mathrm{D_{3/2}}$ fine-structure level.

Under QCDD interrogation, the driven Zeeman sublevels of the $\mathrm{D_{5/2}}$ manifold are symmetrically sampled, resulting in an averaged quadratic shift that can be written as
\begin{equation}
\Delta \nu_{\mathrm{Z2}} = C_{2}\,\langle B^{2} \rangle .
\label{eq:B}
\end{equation}
The quadratic Zeeman coefficient is
\begin{equation}
C_{2} =
\frac{2}{15}
\frac{[\mu_B (g_S - 1)]^{2}}
{h^{2}\nu_{DD}},
\label{eq:C2}
\end{equation}
where $\nu_{DD}$ is the fine-structure splitting between the $\mathrm{D_{5/2}}$ and $\mathrm{D_{3/2}}$ levels \cite{dube2013}.

The mean-squared magnetic field entering Eq.~(\ref{eq:B}) is
\begin{equation}
\langle B^{2} \rangle =
\langle B_{\mathrm{dc}} \rangle^{2}
+
\langle B_{\mathrm{ac}}^{2} \rangle ,
\end{equation}
where $B_{\mathrm{ac}}$ denotes oscillating magnetic fields at the trap-drive frequency.

The amplitude of the oscillating magnetic field is constrained to $\lesssim 40~\mathrm{mG}$ from measurements of modulation indices obtained by comparing Zeeman transitions of opposite magnetic susceptibilities probed along the same laser direction. This procedure separates magnetic contributions from micromotion-induced sidebands \cite{Meir2020Apparatus}.

For ${}^{40}\mathrm{Ca}^{+}$, the quadratic Zeeman coefficient $C_{2}$ has been experimentally measured \cite{zhang2025liquid}. For ${}^{88}\mathrm{Sr}^{+}$, $C_{2}$ is obtained from Eq.~(\ref{eq:C2}) using the known fine-structure splitting. In the absence of an experimental determination for $\mathrm{Sr}^{+}$, the same relative uncertainty is conservatively assigned to both coefficients, as stated in the main text.

We express the fractional quadratic Zeeman shift as
\begin{equation}
\left(\frac{\Delta \nu}{\nu}\right)_{\mathrm{Z2}}
= \left(\frac{C_{2}}{\nu}\right) B^{2},
\end{equation}
with uncertainty
\begin{equation}
u_{\mathrm{Z2}}=
\sqrt{
\left[u\!\left(\frac{C_{2}}{\nu}\right) B^{2}\right]^{2}
+
\left[\frac{C_{2}}{\nu}\,u(B^{2})\right]^{2}
}.
\end{equation}

For $\mathrm{Sr}^{+}$ we use 
$({C_{2}}^{\rm Sr}/\nu)=(7.0198\pm0.007)\times10^{-17}\ \mathrm{G^{-2}}$
and
$B_{\rm Sr}^{2}=9.7718\pm0.0008\ \mathrm{G^{2}}$,
and obtain
\begin{equation}
\left(\frac{\Delta \nu}{\nu}\right)_{\mathrm{Z2}}^{\rm Sr}
=(686.0\pm0.7)\times10^{-18}.
\end{equation}

For $\mathrm{Ca}^{+}$ we use
$({C_{2}}^{\rm Ca}/\nu)=(3.5395\pm0.0035)\times10^{-16}\ \mathrm{G^{-2}}$
and
$B_{\rm Ca}^{2}=6.8719\pm0.0008\ \mathrm{G^{2}}$,
and obtain
\begin{equation}
\left(\frac{\Delta \nu}{\nu}\right)_{\mathrm{Z2}}^{\rm Ca}
=(2432\pm3)\times10^{-18}.
\end{equation}

\section{Interaction with electric fields}

This section provides additional details on the characterization of excess micromotion (EMM) and its associated frequency shifts, which contribute to the second-order Doppler and scalar Stark corrections discussed in the main text.

\subsection{Excess micromotion}

Residual static electric fields displace the ions from the RF null, giving rise to excess micromotion at the trap drive frequency $\Omega_{\mathrm{RF}}$ \cite{Berkeland1998Micromotion}. The resulting phase modulation of the clock laser is quantified using resolved sideband spectroscopy. For a small modulation index $\beta$, the ratio of micromotion sideband and carrier Rabi frequencies is
\begin{equation}
R \equiv
\frac{\Omega_{\mathrm{sb}}^{2}}{\Omega_{\mathrm{carr}}^{2}}
\simeq \frac{\beta^{2}}{4}.
\end{equation}

\begin{figure}[t]
\centering
\includegraphics[trim={0.75cm 7.5cm 1cm 7.5cm},clip,width=\linewidth]{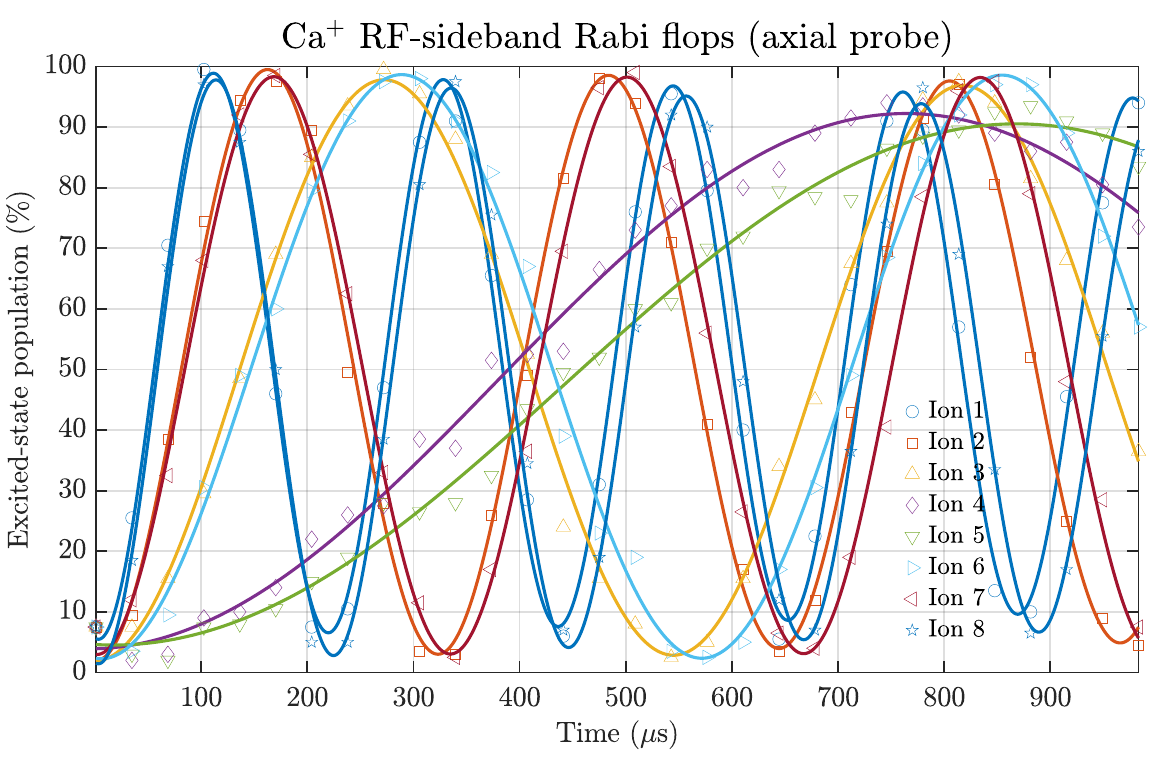}
\caption{Rabi oscillations on the axial micromotion sideband of a ${}^{40}\mathrm{Ca}^{+}$ ion chain. Each trace corresponds to a single ion, illustrating axial variations of the RF field. The carrier Rabi frequency was $\Omega_{\mathrm{carr}}/2\pi= 78$ kHz, and the EMM modulation index varies between $\beta=0.01-0.11$.}
\label{fig:CaAxial Micromotion}
\end{figure}

The modulation index is measured along three non-coplanar probe directions, allowing reconstruction of the full EMM amplitude. Radial EMM can be compensated independently in both radial directions, leaving a dominant axial contribution that varies with ion position along the crystal, as illustrated in Fig.~\ref{fig:CaAxial Micromotion}.

Unbalanced RF currents in the trap electrodes generate an oscillating magnetic field at $\Omega_{\mathrm{RF}}$, which can also modulate magnetically sensitive transitions. The longitudinal component produces an effective phase modulation
\begin{equation}
e^{-i\beta_{B}\cos(\Omega_{\mathrm{RF}} t)}, \qquad
\beta_{B}
= \frac{\chi \mu_{B}}{\hbar \Omega_{\mathrm{RF}}} B_{z,\mathrm{ac}},
\end{equation}
where $\chi = g_{D}m_{D} - g_{S}m_{S}$ is the Zeeman susceptibility~\cite{joshi2024acmagnetic}.

\begin{figure}[t]
\centering
\includegraphics[trim={1.5cm 2cm 1.5cm 2cm},clip,width=\linewidth]{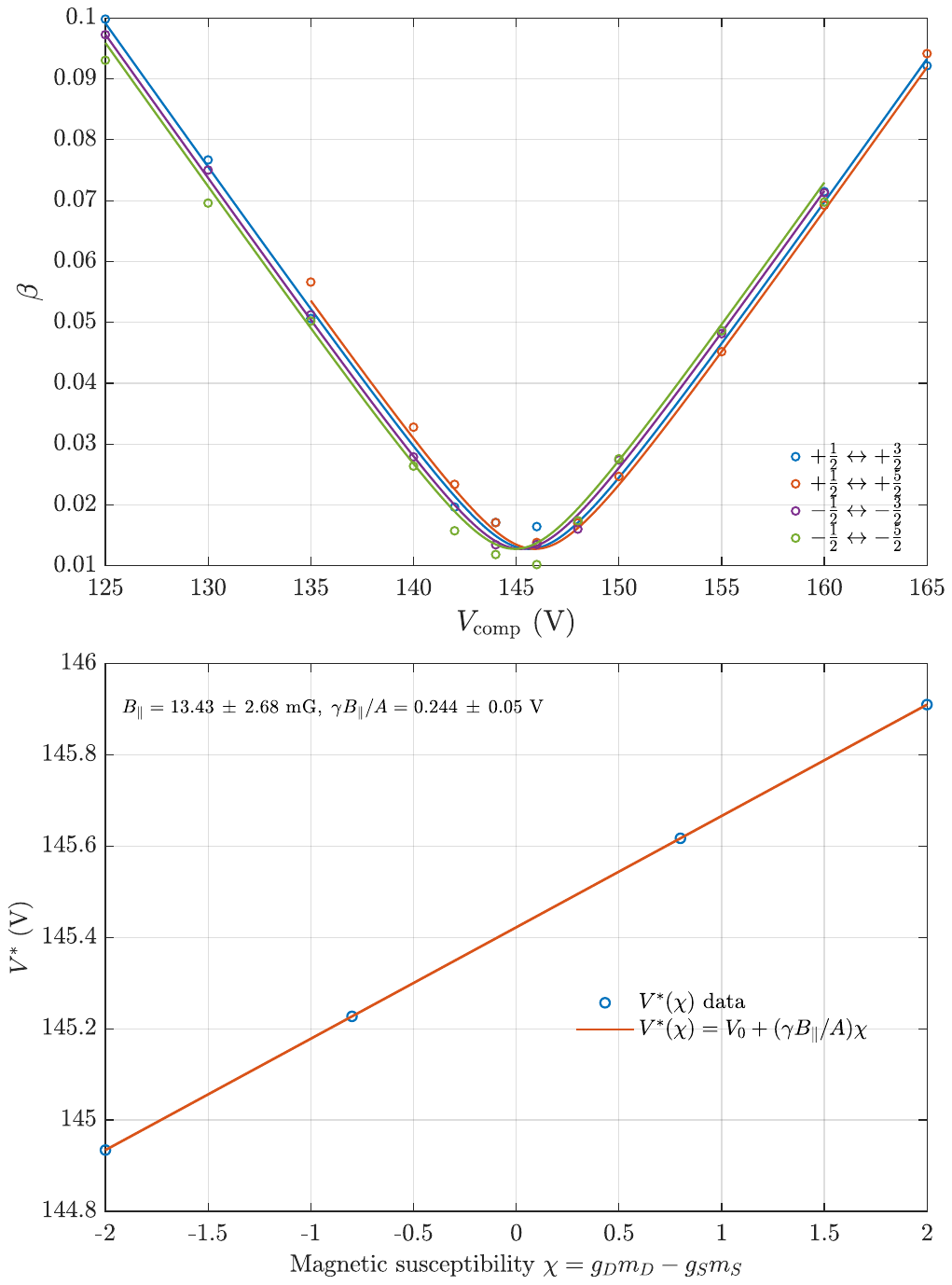}
\caption{Top: Measured modulation index $\beta$ for several Zeeman components of the $S_{1/2}\!\leftrightarrow\!D_{5/2}$ transition in ${}^{88}\mathrm{Sr}^{+}$ together with a fit model describing combined electric and magnetic modulation.
Bottom: Compensation voltage $V^{*}$ as a function of Zeeman susceptibility, used to extract the oscillating magnetic field amplitude $B_{z,\mathrm{ac}}$.}
\label{fig:RF Zeeman}
\end{figure}

By probing transitions with opposite values of $\chi$, electric and magnetic contributions to the measured modulation index can be separated: the average of the modulation indices cancels the magnetic sensitivity, while their difference isolates $B_{z,\mathrm{ac}}$. This procedure is illustrated in Fig.~\ref{fig:RF Zeeman}, where the linear dependence of the compensation voltage on $\chi$ is used to extract an experimental bound on the oscillating magnetic field amplitude. As discussed in Sec.~II, this RF-induced magnetic field contributes to the clock frequency only through the quadratic Zeeman effect.

The RF electric field associated with EMM induces both a relativistic second-order Doppler shift and a scalar quadratic Stark shift. Expressed in terms of the measured sideband ratios $R_i$ along three probe directions,
\begin{align}
\frac{\Delta\nu_{\mathrm{D2,EMM}}}{\nu}
&= -\frac{\langle v_{\mathrm{EMM}}^{2}\rangle}{2c^{2}}
= -\left(\frac{\Omega_{\mathrm{RF}}}{\omega}\right)^{2}
\sum_{i} R_{i}, \\
\frac{\Delta\nu_{\mathrm{Stark,EMM}}}{\nu}
&= -\frac{\Delta\alpha_{0}}{h\nu}
\left(\frac{m \Omega_{\mathrm{RF}}^{2} c}{e \omega}\right)^{2}
\sum_{i} R_{i},
\end{align}
where $\omega$ is the probe laser angular frequency.

The total EMM-induced fractional frequency shift (sum of Doppler and scalar Stark contributions) is
\begin{equation}
\left(\frac{\Delta \nu}{\nu}\right)_{\rm EMM}
= C_{\rm tot,EMM}\,\sum_{i} R_{i},
\end{equation}
\begin{equation*}
C_{\rm tot,EMM} = -\left(\frac{\Omega_{\mathrm{RF}}}{\omega}\right)^{2}\left[1+\frac{\Delta\alpha_{0}}{h\nu}
\left(\frac{mc\Omega_{\mathrm{RF}} }{e}\right)^{2}\right]
\end{equation*}

For $\mathrm{Sr}^{+}$, $\Omega_{\mathrm{RF}}/2\pi=16.15$ MHz, differential polarizability $\Delta\alpha_{0}=-4.8314(20)\times10^{-40} ~\rm Jm^2V^{-2}$ \cite{Lindvall2025Measurement} and the measured total EMM is $\sum_{i} R_{i} = (1.1 \pm 0.5)\times10^{-3}$. Here, we estimate the uncertainty to be half the measured value to account for variation in compensation during data acquisition, as the EMM measurement took place only in between runs. We note that this is a conservative estimation, as the measurements indicate much better stability. We obtain
\begin{equation}
\left(\frac{\Delta \nu}{\nu}\right)_{\rm EMM}^{\rm Sr}
= (0.4 \pm 0.2)\times10^{-18}.
\end{equation}

Similarly, for the $\mathrm{Ca}^{+}$ clock, $\Omega_{\mathrm{RF}}/2\pi=24.85$ MHz, differential polarizability $\Delta\alpha_{0}=-7.2677(21)\times10^{-40} ~\rm Jm^2V^{-2}$ \cite{huang2019ca+} and the measured total EMM is $\sum_{i} R_{i} = (1.5 \pm 0.5)\times10^{-3}$, which gives
\begin{equation}
\left(\frac{\Delta \nu}{\nu}\right)_{\rm EMM}^{\rm Ca}
= (0.15 \pm 0.07)\times10^{-18}.
\end{equation}

\section{Thermal motion}

\begin{figure}[t]
\centering
\includegraphics[trim={0.25cm 8cm 1cm 8cm},clip,width=\linewidth]{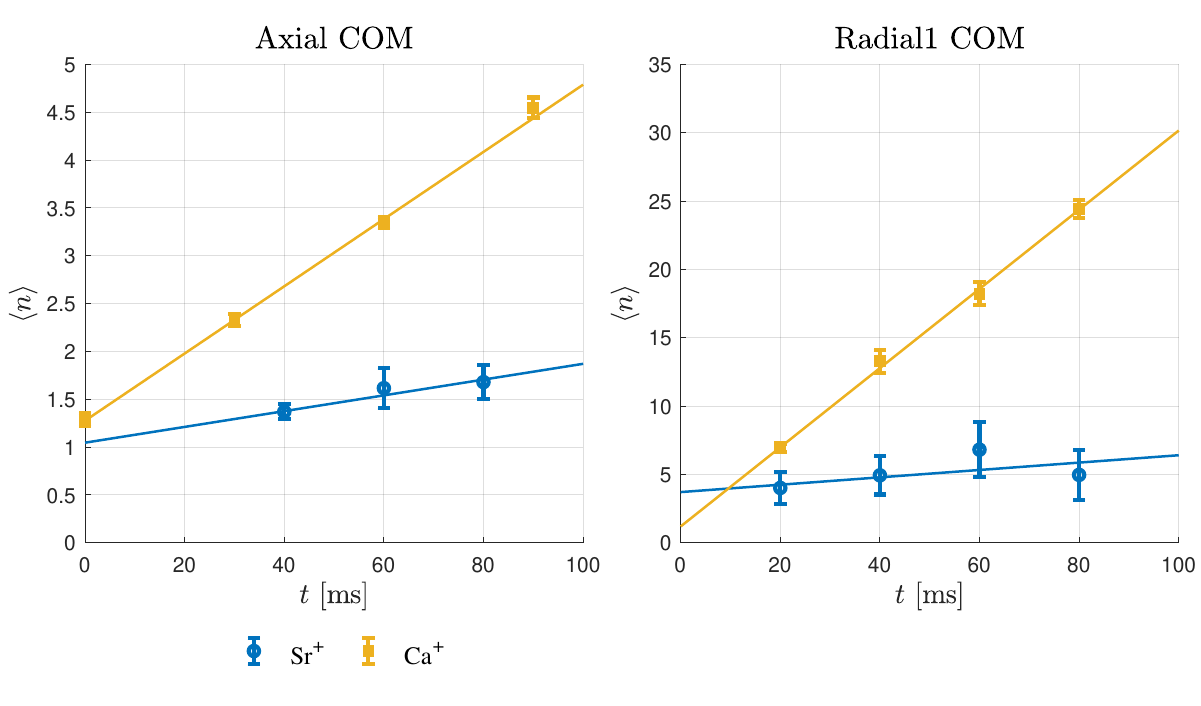}
\caption{Mean phonon occupation of the axial (left) and radial (right) center-of-mass modes of both systems as a function of delay time.}
\label{fig:Heating-Rate}
\end{figure}

Residual thermal motion at the secular frequencies $\omega_i$ gives rise to a second-order Doppler shift. Following Doppler and electromagnetically induced transparency (EIT) cooling, the mean phonon occupations $\bar n_{i}$ are extracted from the dephasing of carrier Rabi oscillations. Heating rates $\dot{\bar n}_i$ are obtained from linear fits to $\bar n_i(t)$ as a function of delay time. Figure~\ref{fig:Heating-Rate} shows the measured phonon occupations and the corresponding linear fits.

Assuming linear heating during the Ramsey free-evolution time $T_R$, the time-averaged phonon occupation is
\begin{equation}
\langle \bar n_i \rangle_{T_R}
= \bar n_{i,0} + \tfrac{1}{2}\dot{\bar n}_i T_R .
\end{equation}
The corresponding mean secular kinetic energy is
\begin{equation}
\langle E_{K} \rangle
= \sum_i \frac{\hbar\omega_i}{2}
\left(\langle \bar n_i \rangle_{T_R} + \tfrac{1}{2}\right),
\end{equation}
leading to the fractional second-order Doppler shift
\begin{equation}
\left(\frac{\Delta \nu}{\nu}\right)_{{\rm D2,sec}}
= C_{{\rm D2}}\,E_{K},
\qquad
C_{{\rm D2}}=-\frac{1}{m c^{2}},
\end{equation}
with uncertainty
\begin{equation}
u_{{\rm D2,sec}}=
\sqrt{
\left(uC_{{\rm D2}}\,E_{K}\right)^{2}
+
\left(C_{{\rm D2}}\,uE_{K}\right)^{2}
}.
\end{equation}

For {$\mathrm{Sr}^{+}$} we use
\[
C_{{\rm D2}}=-7.61\times10^{7}\ \mathrm{J^{-1}},
\]
\[
E_{K}=(1.073\pm0.30)\times10^{-26}\ \mathrm{J},
\]
and obtain
\begin{equation}
\left(\frac{\Delta \nu}{\nu}\right)_{{\rm D2,sec}}^{\rm Sr}
=(-0.82\pm0.23)\times10^{-18}.
\end{equation}

For {$\mathrm{Ca}^{+}$} we use
\[
C_{{\rm D2}}=-1.68\times10^{8}\ \mathrm{J^{-1}},
\]
\[
E_{K}=(2.611\pm0.017)\times10^{-26}\ \mathrm{J},
\]
and obtain
\begin{equation}
\left(\frac{\Delta \nu}{\nu}\right)_{{\rm D2,sec}}^{\rm Ca}
=(-4.37\pm0.03)\times10^{-18}.
\end{equation}

In the pseudopotential approximation, the driven micromotion associated with the trap RF field gives rise to a kinetic energy in each radial mode that is equal to the corresponding secular kinetic energy \cite{Berkeland1998Micromotion}. This driven motion originates from the RF electric field at the ion position and leads to the same physical effects as EMM, namely a relativistic second-order Doppler shift and a scalar quadratic Stark shift. However, the relevant measured quantity here is the radial secular kinetic energy extracted from heating-rate measurements. The total micromotion-induced fractional frequency shift is
\begin{equation}
\left(\frac{\Delta \nu}{\nu}\right)_{\rm mm}
= C_{\rm tot, mm}\,E_{K,{\rm rad}},
\end{equation}
\begin{equation*}
 C_{\rm tot, mm}
= -\frac{1}{m c^{2}}
\left[
1+\frac{\Delta\alpha_{0}}{h\nu}
\left(\frac{mc\Omega_{\mathrm{RF}}}{e}\right)^{2}
\right].
\end{equation*}
The uncertainty is
\begin{equation}
u_{\rm mm}
=
\sqrt{
(uC_{\rm tot, mm}\,E_{K,{\rm rad}})^{2}
+
(C_{\rm tot, mm}\,uE_{K,{\rm rad}})^{2}}.
\end{equation}

We use for {$\mathrm{Sr}^{+}$}
\[
C_{\rm tot, mm}=(2.000\pm0.004)\times10^{7}\ \mathrm{J^{-1}},
\]
\[
E_{K,{\rm rad}}=(1.026\pm0.304)\times10^{-26}\ \mathrm{J},
\]
and obtain
\begin{equation}
\left(\frac{\Delta \nu}{\nu}\right)_{\rm mm}^{\rm Sr}
=
(0.21\pm0.06)\times10^{-18}.
\end{equation}

We use for {$\mathrm{Ca}^{+}$}
\[
C_{\rm tot, mm}=(4.9\pm0.5)\times10^{5}\ \mathrm{J^{-1}},
\]
\[
E_{K,{\rm rad}}=(2.556\pm0.017)\times10^{-26}\ \mathrm{J},
\]
and obtain
\begin{equation}
\left(\frac{\Delta \nu}{\nu}\right)_{\rm mm}^{\rm Ca}
=
(0.0121\pm0.0012)\times10^{-18},
\end{equation}
where the near cancellation reflects an operation close to the magic RF frequency.

\section*{Blackbody radiation (BBR)}

The laboratory temperature is stable to within $0.1~\mathrm{K}$ around
$T_{\rm lab}=295.5~\mathrm{K}$. However, resistive heating of the trap assembly due to RF currents can increase the effective blackbody radiation (BBR) temperature experienced by the ions. We continuously monitor the vacuum-chamber temperature and find it to be stable to 0.1 K as well, however, the value depends on the RF power that is used for operating the ion trap. We measure a temperature susceptibility to RF power of 1.0(1) K/Watt and 0.6(1) K/Watt for the Ca$^+$ and Sr$^+$ vacuum chambers, respectively. Without a direct sensor available for internal trap components, we measured the residual effect by comparing the clocks while operating the Sr$^+$ clock at different RF powers as discussed in the main text. This measurement results in an additional effective temperature susceptibility of $1.3 \pm 1.9$ K/Watt. 

The fractional blackbody radiation shift can be approximated to 
\begin{equation}
\Delta \nu_{\rm BBR}= C_{\rm BBR}\times T_{\rm eff}^{4},
\end{equation}
with uncertainty
\begin{equation}
u\left( \nu_{\rm BBR}\right)=
\sqrt{
\left(uC_{\rm BBR}\,T_{\rm lab}^{4}\right)^{2}
+
\left(C_{\rm BBR}\,uT_{\rm lab}^{4}\right)^{2}
}.
\end{equation}

In all measurements, the  Ca$^+$ trap was operated with RF power of $\rm P=0.56$ Watt, which yields an effective temperature of $297 \pm 1$ K.  With $C_{\rm Ca, BBR}=4.7066\pm0.0003 \times 10^{-11}\ \mathrm{Hz~K^{-4}}$ \cite{huang2022liquid} we obtain  
\begin{equation}
\left(\frac{\Delta \nu}{\nu}\right)_{\rm BBR}^{\rm Ca}
=(891\pm12)\times10^{-18}.
\end{equation}

The Sr$^+$ clock was operated at two different RF powers $\rm P_L=0.4$ Watt and $\rm P_H=1$ Watt,  corresponding to $\ T_{\rm eff,L}=296.2 \pm 0.7$ K and $\ T_{\rm eff,H}=297.4 \pm 1.9$ K.
With $C_{\rm Sr,BBR}=3.0616\pm0.0046 \times10^{-11} \mathrm{Hz~K^{-4}}$ \cite{steinel2023evaluation}
we obtain 
\begin{align}
\left(\frac{\Delta \nu}{\nu}\right)_{\rm BBR,L}^{\rm Sr}
&=529\pm5 \times10^{-18}.\\
\left(\frac{\Delta \nu}{\nu}\right)_{\rm BBR,H}^{\rm Sr}
&=538\pm14 \times10^{-18}.
\end{align}

\section*{Gravitational redshift}

A differential gravitational redshift arises from the vertical separation between the
${}^{88}\mathrm{Sr}^{+}$ and ${}^{40}\mathrm{Ca}^{+}$ clock systems.
The Ca$^{+}$ trap is located higher than the Sr$^{+}$ trap by
\[
\Delta h = (110\pm5)\ \mathrm{mm}.
\]

The corresponding fractional frequency shift is
\begin{equation}
\left(\frac{\Delta \nu}{\nu}\right)_{\rm grav}
=
\frac{g\,\Delta h}{c^{2}},
\end{equation}
where $g$ is the local gravitational acceleration.
Using $g\simeq9.8~\mathrm{m\,s^{-2}}$, we obtain
\begin{equation}
\left(\frac{\Delta \nu}{\nu}\right)_{\rm grav}
=
(1.10\pm0.05)\times10^{-17}.
\end{equation}

Because the Ca$^{+}$ system is located at the higher gravitational potential,
Its clock frequency is increased relative to the Sr$^{+}$ system.
The measured frequency ratio $\nu_{\mathrm{Sr}}/\nu_{\mathrm{Ca}}$ is therefore corrected
by reducing the Ca$^{+}$ frequency by
\[
\Delta\nu_{\rm Ca}
=
(1.10\pm0.05)\times10^{-17}\,\nu_{\rm Ca}.
\]

\section*{Ac Stark (light) shift}
All laser beams are turned off during interrogation by two-stage Switching. One is an AOM in a double-pass configuration. The second is either another single-pass AOM or a MEMS-based voltage control attenuator. We measured the light leakage to the fibers using a photomultiplier and found it to be well below pW. In addition, we ran a comparison without using the second blocking stage and bound the effect to less than 50 mHz. Since the second stage adds another 20-30 dB of attenuation, we can bound light leakage shift to below 0.1mHz.







\bibliographystyle{apsrev4-2}
\bibliography{references}